\documentclass[iicol]{sn-jnl}

\usepackage{graphicx}%
\usepackage{multirow}%
\usepackage{amsmath,amssymb,amsfonts}%
\usepackage{amsthm}%
\usepackage{mathrsfs}%
\usepackage[title]{appendix}%
\usepackage{xcolor}%
\usepackage{textcomp}%
\usepackage{manyfoot}%
\usepackage{booktabs}%
\usepackage{algorithm}%
\usepackage{algorithmicx}%
\usepackage{algpseudocode}%
\usepackage{listings}%

\theoremstyle{thmstyleone}%
\theoremstyle{thmstyletwo}%

\theoremstyle{thmstylethree}%

\begin{document}

\title{Neural Networks for 3D Characterisation of AGATA Crystals}


\author*[1,2]{\fnm{Mojahed} \sur{Abushawish}}\email{mojaheda@ucm.es}

\author[1]{\fnm{Guillaume} \sur{Baulieu}}

\author[1]{\fnm{Jérémie} \sur{Dudouet}}\email{j.dudouet@ip2i.in2p3.fr}

\author[1]{\fnm{Olivier} \sur{Stézowski}}

\affil[1]{\orgdiv{Universite Claude Bernard Lyon 1, CNRS/IN2P3, IP2I Lyon, UMR 5822}, \city{Villeurbanne}, \postcode{F-69100}, \country{France}}
\affil[2]{\orgdiv{Grupo de Fisica Nuclear, Universidad Complutense de Madrid},\street{CEI Moncloa} \city{Madrid}, \postcode{28040}, \country{Spain}}


\abstract{Precise localisation of gamma-ray interactions is crucial for the performance of the Advanced GAmma Tracking Array (AGATA). The Pulse Shape Analysis (PSA) method used for the position estimation of gamma-ray interactions relies on a simulated signal database. The Pulse Shape Comparison Scanning (PSCS) method was used to scan AGATA crystals in order to produce an experimental database of signals.  This paper presents a novel approach using Long Short-Term Memory (LSTM) neural networks to determine the 3D interaction position of gamma rays within AGATA crystals, trained on data from IPHC Strasbourg, allowing for the construction of an experimental database.  A custom masked loss function is introduced to enable training with incomplete position information. The database generated by this new method outperforms the existing simulated database, and the experimental database obtained from the conventional PSCS algorithm.
}




\maketitle

\section{Introduction}
Gamma-ray spectroscopy serves as a powerful technique in experimental nuclear physics, providing insights into nuclear structure and reactions. The Advanced GAmma Tracking Array (AGATA)~\cite{Akkoyun2012,AGATA2} significantly enhances in-beam gamma-ray spectroscopy capabilities. Through the use of highly segmented high-purity germanium (HPGe) detectors, it delivers state-of-the-art energy and position resolution. The precise determination of gamma-ray interaction positions within the detector volume is essential for effective gamma-ray tracking and Doppler correction, ultimately influencing the overall performance of the array.

Current methods for interaction localisation rely on Pulse Shape Analysis (PSA)~\cite{PSA2,PSA3,PSA}. PSA algorithms compare measured signals (traces) from the detector with a pre-calculated basis of signals corresponding to known positions. These bases can be generated through simulations, such as those provided by the AGATA Detector Library (ADL)~\cite{ADL}, or more recently using the AGATAGeFEM package~\cite{AGATAGeFEM}. 

Several techniques have been developed within the collaboration to study the detector response and create experimental bases, called scanning tables. These tables can be divided into three techniques: coincidence scanning~\cite{nelson2007characterisation,ha2013new}, Pulse Shape Comparison Scanning (PSCS) with a collimated $\gamma$ source~\cite{PSCS0}, and PSCS with electronic collimation~\cite{goel2011spatial,hernandez2013characterization}.

The PSCS with a collimated source was first implemented at the Institut pluridisciplinaire Hubert Curien (IPHC)~\cite{PSCS} (Strasbourg scanning table). In this technique, two scans are performed: one with the detector positioned horizontally relative to a collimated source and one positioned vertically. The position of the collimator constrains the position of the $\gamma$-ray interaction in the detector. Horizontal and vertical scans are therefore used to determine the ($X,Z$) and ($X,Y$) positions of the interaction, respectively. The signals from both scans are then compared to build a 3D basis of signals using the PSCS algorithm~\cite{PSCS,decan2020,PSCS2,PSCS3}. A limitation of this approach is that it is computationally intensive and time-consuming, taking approximately $4.7$~days (on a $2.60$~GHz processor
machine) to process data from one scanned crystal~\cite{PSCS2}.

Machine learning (ML) techniques, particularly neural networks (NN), have emerged as powerful tools for data analysis and pattern recognition in various fields, including nuclear physics~\cite{ML_NP}. Recent applications of ML in nuclear instrumentation include particle tracking, event classification, and signal processing~\cite{ML_NP}. The ability of NNs to learn complex, non-linear relationships and patterns directly from the data makes them ideally suited to address the challenges of gamma-ray interaction localisation using detector signals.

The AGATA spectrometer is designed to comprise tapered hexagonal coaxial n-type HPGe crystals, arranged to cover a solid angle of $4\pi$. To achieve this spherical tessellation, the array utilises three distinct asymmetric crystal shapes, designated as types A, B, and C. Despite the slight variations in shape, all crystals share a common length of 90~mm and a rear diameter of 80~mm.

\begin{figure*}[!t]
\centering
\includegraphics[width=0.8\hsize]{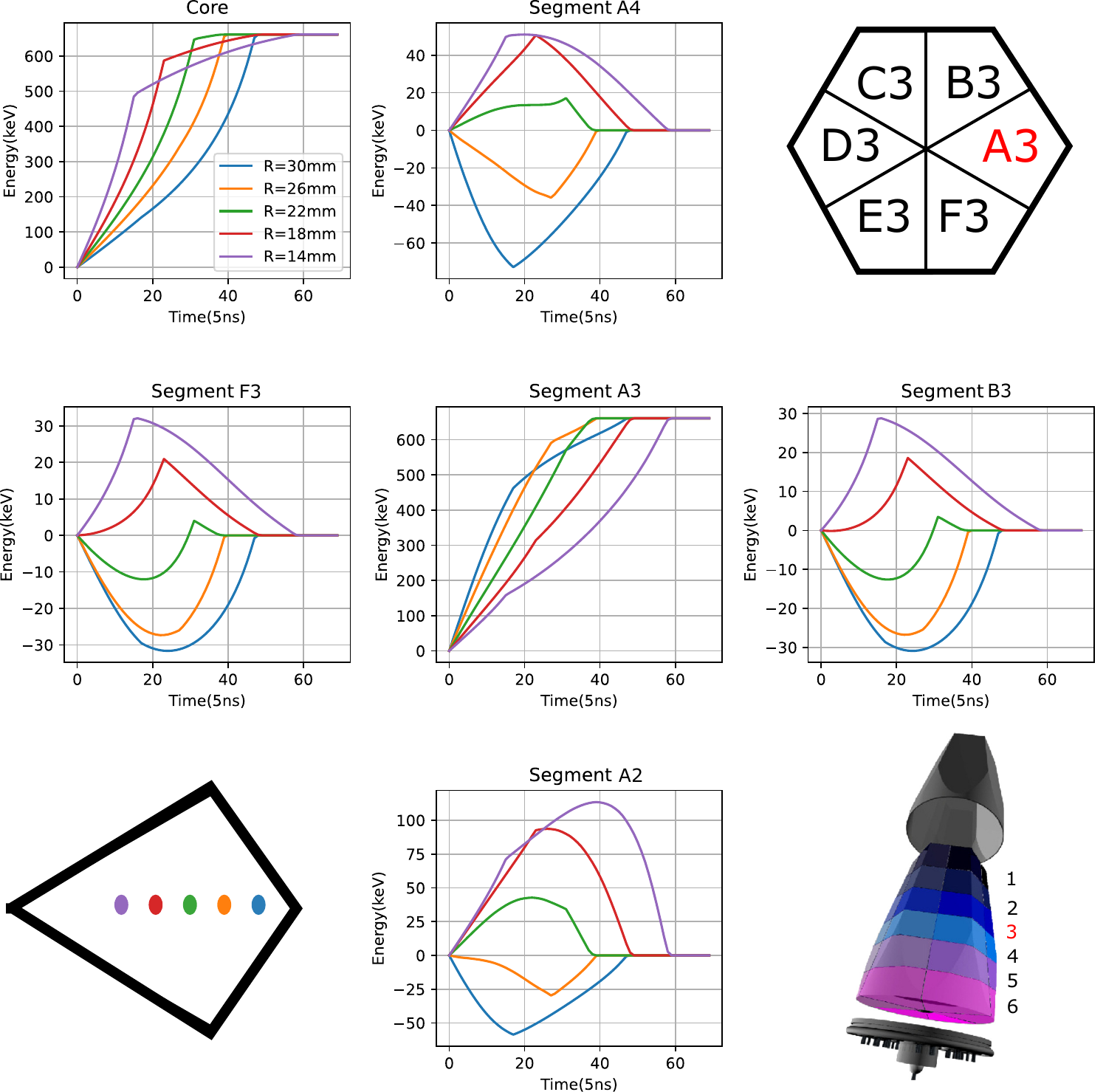}
\caption{An example of ADL simulated AGATA detector signals (traces) for interactions at different radial positions (R) within segment A3. The AGATA nomenclature (top right) and crystal structure (bottom right) are shown for reference. The varying shapes and polarities of the transient signals in neighbouring segments (B3, F3, A2, A4) highlight the position sensitivity used for PSA and NN-based positioning.}
\label{fig:Signals_ADL_R}
\end{figure*}

To enable the precise tracking required for such a setup, each crystal is electrically segmented in 6 angular sectors and 6 depth layers, partitioning the volume into 36 segments. This outer segmentation is complemented by a central inner electrode that serves as the common anode, referred to as the \textit{core}. The nomenclature used to identify these segments is illustrated in Fig.~\ref{fig:Signals_ADL_R}. The hexagonal diagram (top right) shows the six angular sectors (A-F), while the 3D crystal model (bottom right) depicts the six depth layers (1-6) along the z-axis.

Figure~\ref{fig:Signals_ADL_R} illustrates the detector response to simulated gamma-ray interactions using the ADL code. In this context, the signal from a single electrode once treated and filtered is referred to as a \textit{trace}, while the complete set of signals from all 36 segments and the central contact (core trace) is collectively termed a \textit{super-trace}. The figure depicts specific traces for five interactions that occur at different radial positions (R) within a single segment, designated as the \textit{fired segment}. In this example, A3 is the fired segment, referring to the segment in Sector A and Layer 3. The plots display the primary net-charge traces of the core and the fired segment (A3), alongside the transient (or induced) traces from the neighbouring segments (B3, F3, A2, A4). The shapes, amplitudes, and polarities of these transient traces are highly dependent on the interaction's radial and azimuthal position. For example, the transient trace in the outer neighbouring segment (A4) changes from negative to positive as the interaction radius decreases. It is this position-dependent response across the super-trace that forms the basis for position determination.

This work introduces a novel approach to process the Strasbourg scanning table data and produce experimental bases using neural networks. 

\section{Methodology}

Two AGATA crystals were characterised, S001~\cite{decan2020} and A005 using the Strasbourg scanning table. The S001 is a prototype, symmetric AGATA crystal, while the A005 is an asymmetric AGATA crystal (type A) which had been used in the AGATA campaign at GANIL~\cite{GANIL_AGATA}. More details on the AGATA crystals geometries can be found in references~\cite{Akkoyun2012,S001}.

Two 2D scans were performed for each crystal using a $^{137}$Cs source with the 1~mm diameter pinhole collimator with a $2 \times 2 $~mm$^2$ pitch. One with the crystal placed horizontally and one vertically. 

\subsection{PSCS}

The PSCS algorithm is an established technique for generating experimental bases from scanning table data. The PSCS algorithm relies on comparing signals from two intersecting scans (horizontal and vertical). The underlying principle is that signals originating from the same physical interaction point should have similar pulse shapes, regardless of the scan direction. A $\chi^2$ minimisation is employed to quantify the similarity between signals from the two scans using the following formula:

\begin{equation}
\chi^2 = \frac{\sum\limits_{ch=0}^{36} \sum\limits_{i=1}^{120} (V_{ch}(i) - H_{ch}(i))^2}{\sigma^2 \cdot N},
\end{equation}

where $V_{ch}(i)$ and $H_{ch}(i)$ represent the signal amplitudes of channel $ch$ at time sample $i$. The signals are digitised at a sampling frequency of 100~MHz, meaning each time sample corresponds to a 10~ns interval. Here, $\sigma$ represents the noise amplitude, and $N$ is the total number of samples compared (37 channels $\times$ 120 samples). Signal pairs with a $\chi^2$ value below a predefined threshold are considered to originate from the intersection point of the two scans. This process is iterative, refining the selection of signals to create a basis representative of the interaction position.  However, the optimal $\chi^2$ threshold varies depending on the location within the crystal, requiring careful adjustment and complicating the automation of the process. This, combined with the inherently large number of comparisons required for each position, leads to significant computational cost and complexity in the implementation of the PSCS algorithm. 

PSCS-derived bases for the S001 crystal~\cite{decan2020} were provided by collaborators at IPHC Strasbourg for comparison with the neural network-generated bases.

\subsection{Data Preprocessing}
\label{sec:methodo_preprocessing}

Data acquisition was performed using TNT2~\cite{TNT_digi} digital pulse processors. The resulting raw data underwent several pre-processing steps:
\begin{enumerate}
    \item \textbf{Conversion and Compression:} Raw data was converted to the AGATA Data Format (ADF)~\cite{agapro} using the Scanning Table Data Processing (STDPro) software~\cite{scanneddatareader}. This step also compressed the data significantly ($\sim$ factor of 5).
    \item \textbf{Hit Filtering:} Data was filtered to retain only events with a single hit per crystal.
    \item \textbf{Coincidence Filtering:} To filter out random coincidences, events occurring in segments with low statistics were removed. A baseline was established by calculating the average event count from all segments not illuminated by the source. Any event whose fired segment had a total count less than three times this baseline was subsequently removed from the dataset.

    \item \textbf{Timing Alignment:} Core and segment traces were time-aligned using standard AGATA workflow procedures~\cite{stez}.
    \item \textbf{Trace Sampling:} Super-traces were constructed and subsequently reduced in size. From the original 120 samples per trace, only the 60 samples corresponding to the rising edge and plateau were retained, as this window encapsulates the position-sensitive information, and it matches the number of samples used in the standard AGATA workflow. This yields a final super-trace dimension of $(36 \text{ segments} + 1 \text{ core}) \times 60 \text{ samples} = 37 \times 60 = 2220$ features per event.
    \item \textbf{Format Conversion:} Data was converted from ROOT format to NumPy arrays using the `uproot' library~\cite{uproot} for compatibility with machine learning frameworks.
    \item \textbf{Dataset Preparation:} The datasets from the vertical and horizontal scans were combined, and the event sequence was randomly permuted to ensure a homogeneous mixture of data points. The combined dataset was saved in HDF5 format which allows for loading specific events from the files without the need to load the entire file into memory, a feature that becomes extremely important when building the experimental basis.
    \item \textbf{Train/Validation Split:} The full dataset for each crystal (around $10^8$ events per crystal) was split into a training set (90\%) and a validation set (10\%) to monitor model performance and prevent overfitting.
\end{enumerate}

The training dataset for the NN model consisted of events within the energy ranges of 300-480 keV (upper Compton continuum) and 650-670 keV (photopeak region). This selection was based on a preliminary analysis indicating that these energy windows correspond to the lowest prediction errors achievable by the model. By concentrating the training on these high-fidelity regions, the aim was to optimise the model's ability to learn the position-dependent features from the most informative signals, while reducing the impact of potentially less reliable data from lower energies or complex scattering regions known to exhibit higher errors and potentially more background contamination.

\subsection{Machine Learning Approach}
\label{sec:methodo_ml}

A NN model was developed to predict the 3D interaction position $(\hat{x}, \hat{y}, \hat{z})$ from the input super-trace. However, the scanning data provides only partial ground truth, meaning one coordinate is always unknown during training. To address this, we employed a specific strategy utilising a masked loss function. In machine learning, the loss function calculates the error between the model's prediction and the actual target; by masking this function, we allow the model to learn exclusively from the available dimensions, ignoring the error on the unknown coordinate.
\subsubsection{Masked Loss Function}
\label{sec:methodo_maskedloss}

Since the ground truth provides only two spatial coordinates for each super-trace (X, Y for vertical; X, Z for horizontal), a standard 3D loss function cannot be directly applied. A masked loss function was implemented to compute the error only along the known dimensions for each training sample. The masked Euclidean distance loss is defined as:
\begin{equation}
L = \frac{1}{N} \sum_{k=1}^{N} \sqrt{
    \begin{gathered} 
        \left[ M_x (x_k - \hat{x}_k)^2 \right. \\ 
        \left. + M_y (y_k - \hat{y}_k)^2 + M_z (z_k - \hat{z}_k)^2 \right]
    \end{gathered}
}
\label{eq:masked_euclidean}
\end{equation}

where $(x_k, y_k, z_k)$ is the true scanned position, $(\hat{x}_k, \hat{y}_k, \hat{z}_k)$ is the predicted position, $N$ is the data samples, and $M_x, M_y, M_z$ are mask values (0 or 1). For a vertical scan sample, $M_x=1, M_y=1, M_z=0$. For a horizontal scan sample, $M_x=1, M_y=0, M_z=1$. The mask is applied only during training; during inference, the model predicts all three coordinates. The Euclidean distance loss function was used for its robustness against potential outliers. This implementation maintains the continuity of the loss function, which is a prerequisite for training the model via backpropagation.

\subsubsection{Network Architecture}
\label{sec:methodo_architecture}

\begin{figure}[!ht]
\centering
\includegraphics[width=1\hsize]{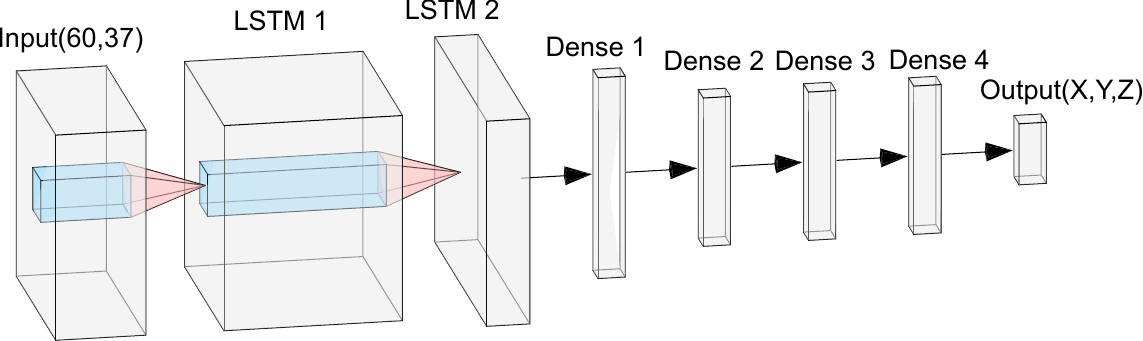} 
\caption{Schematic diagram illustrating the NN architecture.}
\label{fig:NN_archi_methodo}
\end{figure}

The NN architecture employed is based on Long Short-Term Memory (LSTM) layers~\cite{LSTM}. As the detector signals (super-traces) represent time-sequential data where the temporal evolution encodes position information, LSTMs are particularly suitable. They are a type of Recurrent Neural Network (RNN) specifically designed to effectively learn and remember long-term dependencies within sequences. This capability is crucial for capturing the subtle features across the entire pulse shape relevant for accurate position determination. Moreover, LSTMs has been shown to be robust to time misalignment~\cite{FABIAN}, which is crucial to successfully process AGATA signals. These LSTM layers are followed by dense layers for the final position regression. The input to the model is the super-trace, formatted as a single continuous vector of 2220 data points, the architecture includes an initial reshaping step to recover the temporal structure required by the LSTM. The detailed architecture, illustrated schematically in Fig.~\ref{fig:NN_archi_methodo}, consists of the following layers:

\begin{enumerate}
    \item Input Layer: Takes the flattened super-trace (2220 features).
    \item Gaussian Noise Layer: Adds noise to the input signals (see Section~\ref{sec:methodo_overfitting}).
    \item Reshape Layer: Reshapes the input into a $(37 \text{ segments}, 60 \text{ samples})$ format.
    \item Permute Layer: Permutes dimensions to $(60 \text{ samples}, 37 \text{ segments})$ to treat samples as the time sequence.
    \item LSTM Layer 1: 500 memory state size, returns sequences (output shape: (60, 500)).
    \item LSTM Layer 2: 500 memory state size, does not return sequences (output shape: (500)).
    \item Dense Layers: Four fully connected layers with 200, 100, 100, and 100 nodes respectively, using the Rectified Linear Unit (ReLU) activation function.
    \item Output Layer: A dense layer with 3 output nodes (for $\hat{x}, \hat{y}, \hat{z}$) using a linear activation function.
\end{enumerate}

A systematic study of the network architecture indicated that increasing the number or size of dense layers did not improve accuracy, while the number of LSTM units affected convergence speed but not final accuracy if training iterations were adjusted proportionally.

\subsubsection{Training Paradigm}
\label{sec:methodo_training}

The models were trained using the Adam optimiser~\cite{kingma2014adam} with a learning rate of $10^{-4}$. A batch size of 500 super-traces per iteration was used, with a total of $3 \times 10^4$ iterations per model. The super-traces were normalised by scaling each one so that its peak amplitude equaled 1000 arbitrary units. This normalisation procedure relies on the standard assumption that the fundamental shape of a signal is independent of the total energy deposited. This approach is consistent with the established AGATA PSA methodology. The masked Euclidean distance (Eq.~\ref{eq:masked_euclidean}) served as the primary loss function. Throughout this work, the ``error" is defined as the loss value, unless otherwise specified.

Initial attempts to train a single model for the entire crystal revealed issues related to the non-uniform distribution of training data across segments and positions. Segments or positions with significantly more data tended to dominate the loss calculation, leading to biased performance. Several strategies were tested:
\begin{itemize}
    \item \textbf{Training on a uniformly sampled subset} (equal number of super-traces per scanned point): This drastically reduced the dataset size (to ~2\% of the original) and resulted in poorer overall performance~\cite{jrjc}.
    \item \textbf{Training with balanced batches} (equal number of super-traces per segment per batch): This showed improvement over the uniform subset but was still suboptimal.
    \item \textbf{Training one model per segment:} This approach proved most effective. Separate datasets were created for each of the 36 segments, and an independent NN model (with the architecture described above) was trained for each segment, using only the data where that segment is the fired one.
\end{itemize}

\begin{figure}[!ht]
\centering
\includegraphics[width=1.0\hsize]{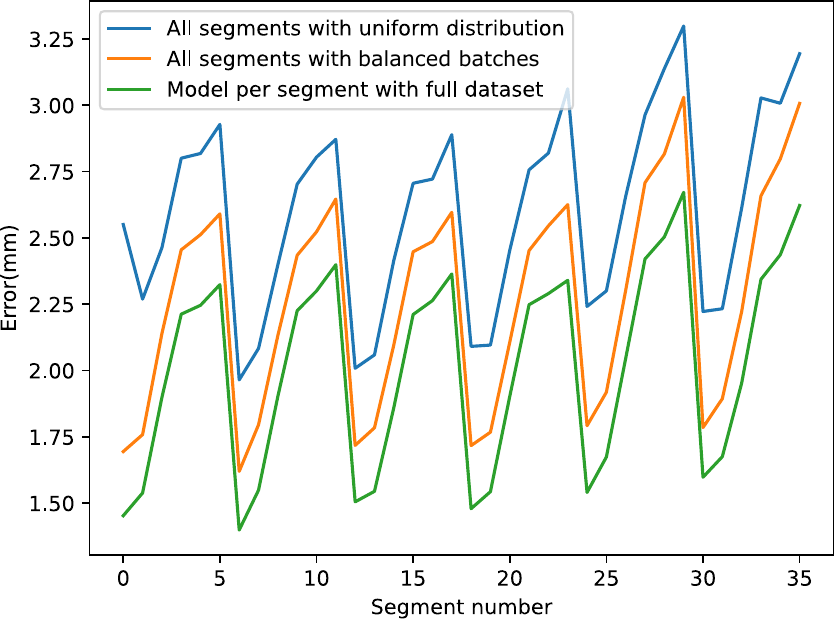}
\caption{The error (mm) on the validation dataset at photopeak energy versus segment number for different training methods: using a uniformly sampled subset (blue), using the full dataset with a single model trained with balanced batches per segment (orange), and training one model per segment (green). Error is calculated on the two known axes. The segments index is explained in the text.}
\label{fig:Training_process_error_methodo}
\end{figure}

Figure~\ref{fig:Training_process_error_methodo} displays the prediction error calculated on the two known axes plotted against the segment index (0--35), where segments 0-5 correspond to Sector A (Layers 1-6), segments 6-11 correspond to Sector B (Layers 1-6), and so on, up to segments 30-35 for Sector F. Two distinct patterns are observable in the error distribution. First, the error is significantly higher for the rear segments (Layers 5 and 6) across all sectors. This is attributed to the larger physical volume of these segments, as well as reduced statistics caused by beam attenuation during the vertical scan (front entry). Second, a general increase in error is observed in Sectors E and F; this arises from attenuation during the horizontal scan. In this configuration, the detector was oriented with Sectors B and C facing the source. Consequently, the beam was attenuated by the crystal bulk before reaching the opposing sectors (E and F), resulting in lower statistics. Despite these physical variations, the per-segment training approach yielded the lowest average error and mitigated the distribution bias observed with a single global model. All subsequent results are based on this per-segment training strategy.

\subsubsection{Overfitting Mitigation}
\label{sec:methodo_overfitting}

\begin{figure}[!ht]
\centering
\includegraphics[width=1.0\hsize]{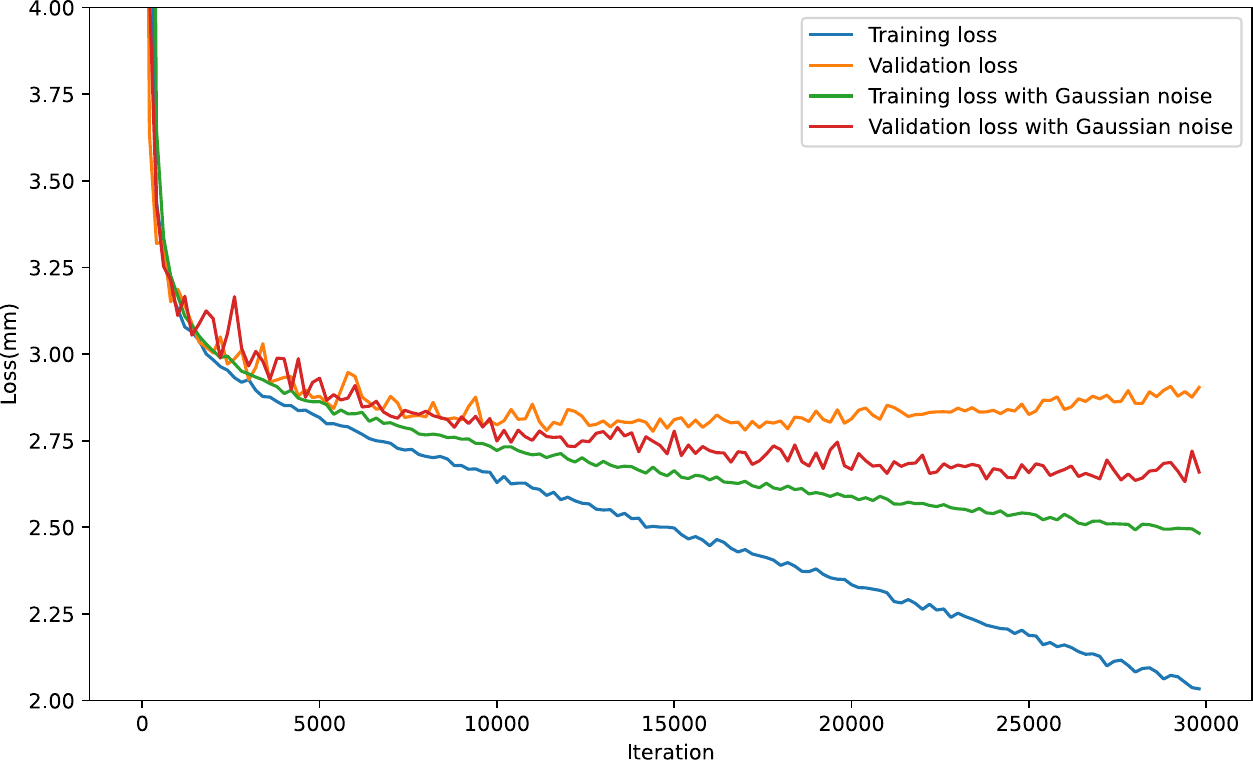}
\caption{Training and validation loss curves for segment F6, without (blue/orange) and with (green/red) the Gaussian noise layer, demonstrating reduced overfitting.}
\label{fig:Overfitting_loss_methodo}
\end{figure}

As the training of the model progressed, it started to exhibit signs of overfitting. Figure~\ref{fig:Overfitting_loss_methodo} tracks the evolution of the loss function throughout the training process, plotted against the number of iterations. The blue curve, representing the training loss, shows a continuous decline, indicating that the model is successfully minimising error on the data it is actively learning from. In contrast, the validation loss (orange curve), calculated on data the model has never seen, initially decreases but eventually reaches a minimum before beginning to rise. This divergence indicates that the model has started to memorise the specific noise of the training set rather than learning generalisable features, leading to degraded performance on new data.

To combat this overfitting and encourage generalisation, a Gaussian noise layer was added after the input layer in the network architecture. This layer adds random noise, drawn from a Gaussian distribution, to each sample of the input super-trace during training. Various noise levels (standard deviations) were experimentally tested during model development. While higher noise levels could further reduce the discrepancy between training and validation loss, they also tended to degrade the overall prediction accuracy (increase the minimum validation loss). Conversely, lower noise levels were less effective at controlling the overfitting. A standard deviation of 5 keV was ultimately selected as it provided the best observed balance between these competing factors. As shown in Fig.~\ref{fig:Overfitting_loss_methodo}, the addition of this 5 keV noise significantly reduced the gap between training and validation loss (green and red curves, respectively) compared to training without noise, and more importantly, led to a lower converged validation loss, indicating improved model generalisation.

\subsection{Evaluation Using Mean Standard Deviation}
\label{sec:methodo_meanstd}

To assess the consistency of the signals corresponding to the predicted positions (since the full 3D ground truth is unavailable), a metric termed Mean Standard Deviation (MSD) was used. For a given predicted 3D voxel position, the following procedure has been applied:
\begin{enumerate}
    \item Collect all super-traces (from the validation set or full dataset inference) predicted by the NN model to be at that position. For comparison, super-traces derived from the PSCS algorithm were also analysed.
    \item Normalise the energy scale of all collected super-traces (scaling the maximum of the signal to 1000 units).
    \item For each time sample $j$ in each trace $i$ (core and neighbours), calculate the standard deviation $\sigma_{ij}$ across all $N$ super-traces assigned to that voxel.
    \item Compute the mean of these standard deviations over all relevant samples and traces (specifically, the first neighbours and the core): $\text{MSD} = \frac{1}{N_{samples} \times N_{traces}} \sum_{i,j} \sigma_{ij}$.
\end{enumerate}

A lower MSD indicates higher consistency among the super-traces predicted at the same position. This metric was used both to evaluate the overall quality of the NN predictions compared to PSCS and as part of an iterative filtering process to remove outlier signals before generating the final experimental bases. The filtering involved repeatedly calculating the MSD, identifying and removing the super-traces most different from the mean super-trace at that position, and stopping when the improvement in MSD plateaued (monitored via its second derivative). Similar filtering method was applied to the PSCS signals~\cite{PSCS}.

\subsection{Validation using Emulated 1-D Scans}
\label{sec:methodo_1dscan}

A key question is whether the model, trained with only 2 known coordinates per trace, can accurately predict the third, unknown coordinate. Since direct validation of the third coordinate is impossible with the 2D scanning data, a test was performed using simulated 1D scans derived from the existing 2D data.
The methodology was as follows:
\begin{enumerate}
    \item Create a 1D dataset:
        \begin{itemize}
            \item Take the vertical scan data (X, and Y known) and create two subsets: one retaining only the X label, the other retaining only the Y label.
            \item Take the horizontal scan data (X, and Z known) and create two subsets: one retaining only the X label, the other retaining only the Z label.
            \item Combine these four subsets into a single training dataset where each super-trace has only one known coordinate associated with it.
        \end{itemize}
    \item Train a NN model (using the same architecture and per-segment strategy as described in section~\ref{sec:methodo_architecture}) on this 1D dataset, using a mask that allows loss calculation only on the single known axis, Ex: ($M_x=1, M_y=0, M_z=0$ if X is known).
    \item Evaluate this 1D-trained model on the validation set: calculate the prediction error for the axes that were unknown during its training (for a trace originally from the vertical scan where only X was known during 1D training, evaluate the error on the predicted Y coordinate using the true Y value from the original 2D data).
    \item Compare this error on the unknown axes from the 1D training with the error on the same axes obtained from the standard model trained on the 2D dataset.
\end{enumerate}

\begin{figure}[!htb]
\centering
\includegraphics[width=1.0\hsize]{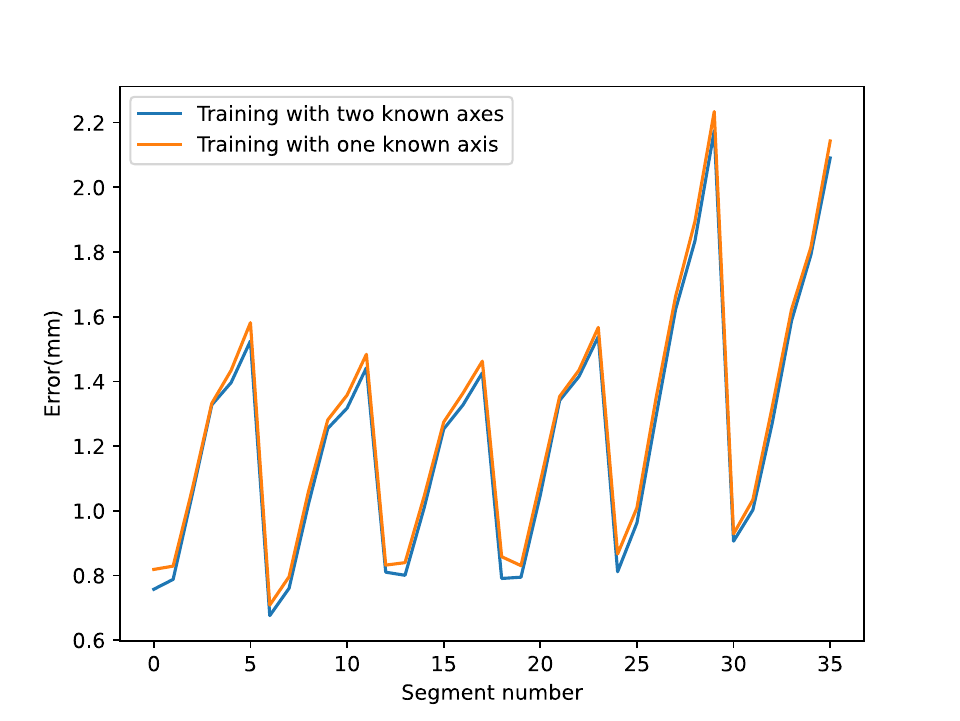} 
\caption{The mean absolute error on the validation dataset versus segment number. Comparison between a model trained on the standard 2D dataset (blue) and a model trained on the simulated 1D dataset (orange). The error shown is calculated only for the axis that was unknown during the 1D training but known in the original 2D dataset. The segment indexing follows the same nomenclature as defined for Fig.~\ref{fig:Training_process_error_methodo}.}
\label{fig:1D_error_methodo}
\end{figure}

Figure~\ref{fig:1D_error_methodo} compares the results of the two methods, presenting the mean absolute error on the previously unknown axis for the 1D-trained model, the same error patterns are observed as in Fig.~\ref{fig:Training_process_error_methodo}. The error for the 1D-trained model is only marginally higher than the error for the 2D-trained model. This suggests that the model learns the underlying relationship between pulse shape and 3D position effectively, and is capable of generalising to predict the unknown coordinate(s) with reasonable accuracy, validating the approach of training with partial ground truth for 3D position prediction. This test provides confidence that training on two 2D scans can yield a model capable of full 3D prediction, potentially enabling future, faster 1D scanning campaigns. 

\subsection{Implementation Details}
\label{sec:methodo_implementation}

All code for data processing, model training, and analysis was written in Python 3. The neural networks were implemented and trained using the TensorFlow library~\cite{abadi2016tensorflow} with the Keras API~\cite{keras2015-library}. Training was performed on a dedicated machine learning server at IP2I, utilising NVIDIA RTX6000 GPUs.

\section{Results}
\label{sec:results}

The performance of the NN model is evaluated based on hit distributions, signal consistency (measured by MSD (see section \ref{sec:methodo_meanstd}.)), and by comparing PSA results obtained using different signal bases. This section will primarily present the results for the A005 crystal. We will only use S001 crystal for the comparison of signal consistency (MSD) between the NN and PSCS algorithm as PSCS bases were not generated for the A005 crystal.

\subsection{Hit Distributions and Error Energy Dependence}
\label{sec:results_hitdist}

The NN model, trained independently for each segment, predicts the 3D interaction position for each input super-trace. The overall performance metrics indicate successful training with minimal overfitting. The mean error on the validation dataset, calculated on the two known axes, was 2.00~mm for the S001 crystal and 1.88~mm for the A005 crystal. The slightly lower error for A005 may be attributed to geometrical differences affecting sensitivity or variations in data quality.

\begin{figure}[!htb]
\includegraphics[width=1\hsize]{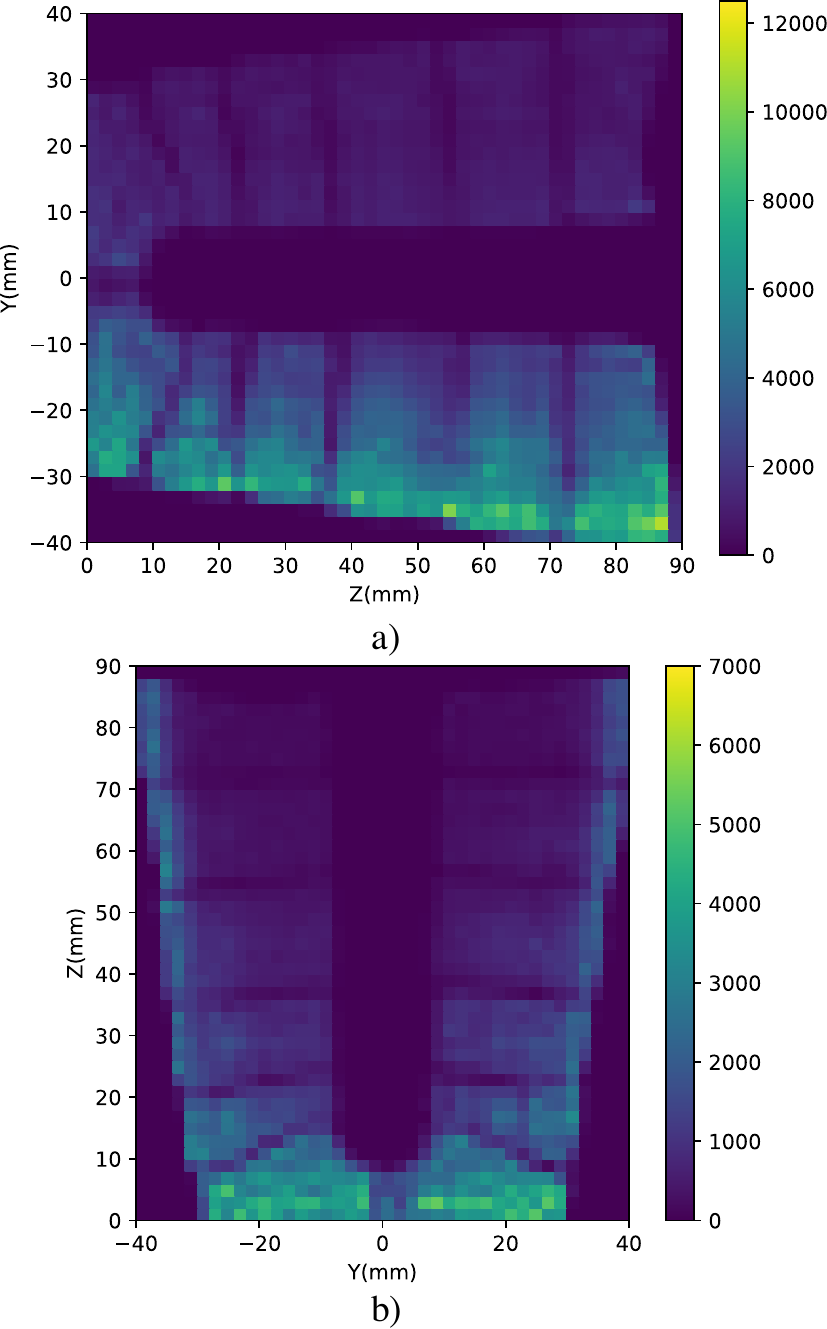}
\caption{Predicted interaction positions in the (Y, Z) plane for X between -3 and 3~mm for the A005 crystal, showing the attenuation effect for horizontal (a) and vertical (b) scans.}
\label{fig:results_dist_s001}
\end{figure}

Figure~\ref{fig:results_dist_s001} shows the distribution of predicted interaction points in the (Y, Z) plane for a central slice (-3 mm $<$ X $<$ 3 mm) for the A005 crystal, for the horizontal scan (a) and for the vertical scan (b). The distributions correctly reproduce the expected attenuation effect of gamma rays traversing the crystal material, with higher hit density closer to the entry surface for each scan type.

\begin{figure}[!htb]
\centering
\includegraphics[width=1\hsize]{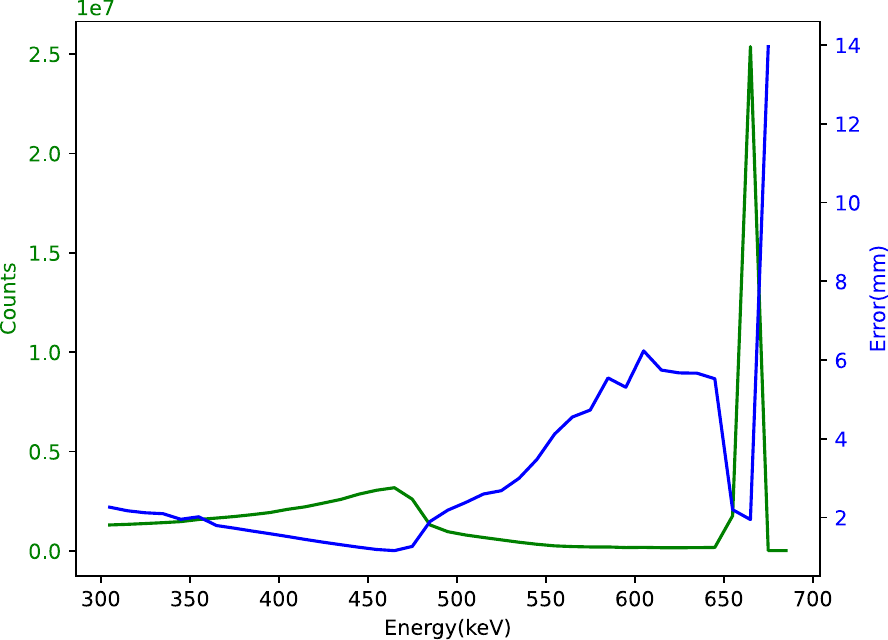}
\caption{Prediction error (blue, right axis) and event statistics (green, left axis) as a function of the hit energy for the A005 crystal.}
\label{fig:results_energy_A005}
\end{figure}

Figure~\ref{fig:results_energy_A005} displays both the prediction error (blue, right axis) and the event statistics (green, left axis) as a function of the deposited energy for the A005 crystal. Compton events (300-480~keV) generally exhibit lower error than photopeak events (around 662 keV). This is consistent with previous findings~\cite{PSCS} and attributed to the fact that photopeak events often involve multiple interactions within the same segment, leading to an averaged pulse shape whose effective position might deviate from the initial interaction point targeted by the scan. The error decreases for higher energy Compton events, possibly due to a reduced probability of random coincidences. The regions between the Compton edge (477 keV) and the photopeak, and above the photopeak, show high error and very low statistics, as single interactions in these energy ranges are physically unlikely.

\subsection{Signal Consistency: Mean Standard Deviation (S001)}
\label{sec:results_meanstd}

Evaluating the quality of the predicted 3D positions is challenging due to the lack of full 3D ground truth. One measure of quality is the consistency of the super-traces assigned to the same predicted voxel. The MSD metric quantifies this consistency. The lower the MSD value is the more similar are the super-traces to each other at a given predicted position. The PSCS dataset included only the photo-peak events, therefore only these events were evaluated.

\begin{figure}[!htb]
\includegraphics[width=1\hsize]{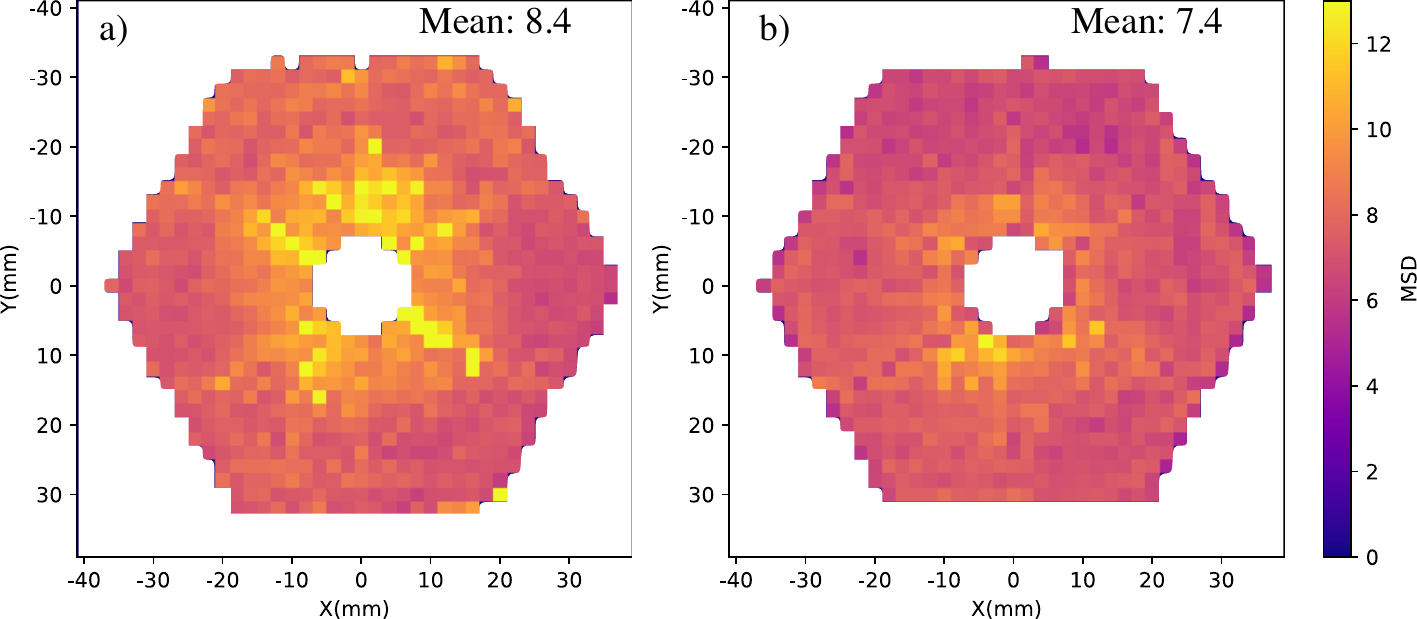}
\caption{MSD per pixel at Z=30~mm for the S001 crystal, comparing the PSCS algorithm results (a) and the NN model predictions (b). Lower values indicate higher signal consistency.}
\label{fig:results_meanstd_comp}
\end{figure}

The MSD was calculated for all predicted positions. When averaged over the entire crystal volume, the MSD for the NN model is 7.8~a.u, a significant improvement over the 10.6~a.u value obtained for the PSCS algorithm. This global trend of superior consistency is illustrated visually in Fig~\ref{fig:results_meanstd_comp}, which compares the MSD per voxel for a representative slice of the S001 crystal at Z=30 mm.
The map generated from the NN model (b) exhibits markedly higher homogeneity and fewer localised regions of high deviation (hot spots) compared to the map from the PSCS algorithm (a). This visual assessment is confirmed quantitatively for this specific slice, where the mean MSD was 7.4~a.u for the NN model versus 8.4~a.u for PSCS. These results, consistent with the trend observed throughout the entire crystal volume, indicate that the NN method produces a more uniform and reliable set of signal bases.

\subsection{Experimental Signal Bases (A005)}
\label{sec:results_signals}

The primary goal of applying the NN model to the scanning data is to generate high-quality experimental signal bases for use in PSA. These bases are constructed by averaging the filtered super-traces predicted at each voxel position. Filtering included selecting energy ranges with minimal error (300-480 keV and 650-670 keV, based on Fig.~\ref{fig:results_energy_A005}), retaining events with prediction error $<$ 2 mm relative to the scanned position, and applying the iterative MSD filtering described above. This resulted in a basis set containing signals for 49,220 voxels.

\subsection{Pulse Shape Analysis Performance}
\label{sec:results_psa}

The quality of the generated NN experimental bases (NN bases) was evaluated by using them within the standard AGATA PSA framework~\cite{stez} (full grid search, 300-700 keV energy range) to predict the interaction positions of the scanned super-traces. The performance was compared against PSA using bases derived from the standard simulated ADL bases and simulated bases from AGATAGeFEM.

The position resolution (FWHM) achieved with the three different signal bases for the two known axes of each scan is summarised in Table~\ref{tab:psa_errors}. The experimental NN bases (3.0-3.1~mm) clearly demonstrate superior performance, yielding the best resolution overall. Among the simulated bases, the AGATAGeFEM bases (4.8-6.2~mm) surpassed the performance of the ADL bases(5.2-8.1~mm), a result consistent with previous study~\cite{AGATAGeFEM}.

A more detailed analysis reveals interesting anisotropies in the position resolution, particularly for the simulated bases. For the vertical scan, the NN basis was perfectly isotropic with $FWHM_x = FWHM_y = 3.1$~mm. The AGATAGeFEM basis was also nearly isotropic ($FWHM_x = 6.0$~mm vs. $FWHM_y = 6.2$~mm), while the ADL basis showed a more notable discrepancy, with its $FWHM_y$ being 0.5~mm larger than its $FWHM_x$ (8.1~mm vs. 7.6~mm).

A more significant trend was observed in the horizontal scan data, where both simulated bases exhibited markedly better resolution along the depth ($z$) axis compared to the transverse ($x$) axis. For AGATAGeFEM, the $FWHM_z$ was 4.8~mm, an improvement of over 1.3~mm compared to its $FWHM_x$ of 6.1~mm. This effect was even more pronounced for the ADL basis, with a resolution of 5.2~mm in $z$ versus 7.8~mm in $x$. In contrast, the NN basis maintained its highly isotropic performance in the X-Z plane, with comparable FWHM values of 3.1~mm and 3.0~mm, respectively.

\begin{table*}[ht!]
\centering
\caption{Comparison of 3D position resolution (FWHM, mm) for the A005 crystal from two scan geometries. Resolution is calculated against the original scan position and the NN model's prediction using different PSA bases. A dash (-) indicates a metric is not applicable for that scan's reference.}
\label{tab:psa_errors}
\begin{tabular}{l ccc ccc}
\toprule
& \multicolumn{6}{c}{\textbf{Position Resolution (FWHM, mm)}} \\
\cmidrule(lr){2-7}
& \multicolumn{3}{c}{vs. Scan Position} & \multicolumn{3}{c}{vs. NN Model} \\
\cmidrule(lr){2-4} \cmidrule(lr){5-7}
\textbf{PSA Basis} & \textbf{$FWHM_x$} & \textbf{$FWHM_y$} & \textbf{$FWHM_z$} & \textbf{$FWHM_x$} & \textbf{$FWHM_y$} & \textbf{$FWHM_z$} \\
\midrule
\multicolumn{7}{l}{\textit{Vertical Scan}} \\
\midrule
NN         & 3.1 & 3.1 & -    & 2.5 & 2.4 & 2.2 \\
AGATAGeFEM & 6.0 & 6.2 & -    & 5.4 & 5.5 & 4.1 \\
ADL        & 7.6 & 8.1 & -    & 7.0 & 7.4 & 4.2 \\
\midrule
\multicolumn{7}{l}{\textit{Horizontal Scan}} \\
\midrule
NN         & 3.1 & -    & 3.0 & 2.4 & 2.5 & 2.2 \\
AGATAGeFEM & 6.1 & -    & 4.8 & 5.5 & 5.1 & 4.1 \\
ADL        & 7.8 & -    & 5.2 & 7.1 & 6.2 & 4.5 \\
\bottomrule
\end{tabular}
\end{table*}

\begin{figure*}[!t]
\centering
\includegraphics[width=1\hsize]{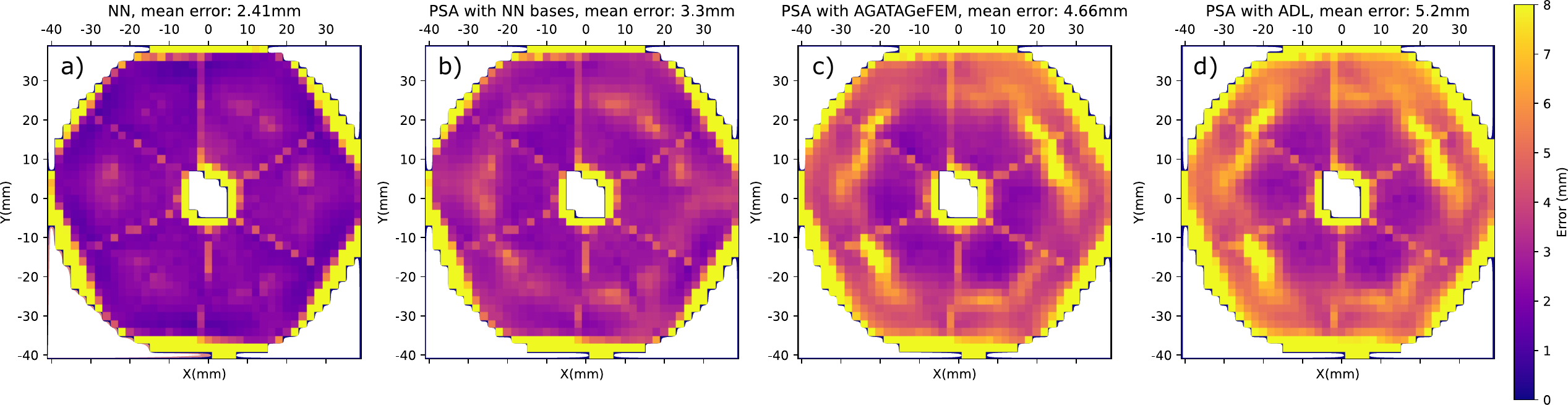}
\caption{PSA position error map for the A005 vertical scan across the fourth layer of the crystal. Comparison of the direct NN model prediction (a) and PSA results using NN bases (b), AGATAGeFEM bases (c), and ADL bases (d). Yellow colour is above 8~mm.}
\label{fig:results_psa_errorv_a005_L4}
\end{figure*}

\begin{figure*}[!t]
\centering

\includegraphics[width=1\hsize]{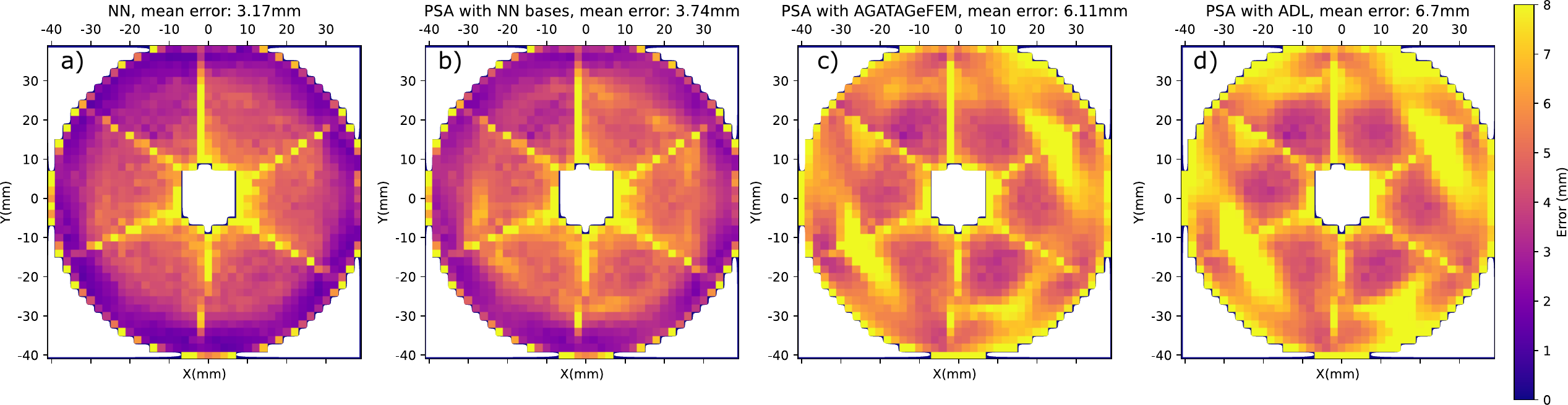} 
\caption{PSA position error map for the A005 vertical scan across the sixth (back) layer of the crystal. The panel arrangement is identical to Fig.~\ref{fig:results_psa_errorv_a005_L4}. Yellow colour is above 8~mm.}
\label{fig:results_psa_errorv_a005_L6}
\end{figure*}

Figures~\ref{fig:results_psa_errorv_a005_L4} and~\ref{fig:results_psa_errorv_a005_L6} show the PSA error maps for the A005 vertical scan across two different depths: the fourth layer and the back-most sixth layer, respectively. This comparison highlights both consistent performance trends and the impact of interaction depth.
Across both layers, several key observations hold true. The direct prediction from the NN model itself (panel (a) in both figures) consistently provides the lowest apparent error and the best homogeneity. Among the PSA-based results, the experimental NN bases (panel b) yield a superior performance compared to the simulated AGATAGeFEM and ADL bases (panels c and d, respectively). The simulated bases, in particular, show degraded performance at radii $\geq$ 20 mm, suggesting potential issues in accurately simulating the core signal rise time in the outer regions of the crystal.
A region of increased error is evident in the middle of the segments as it can be seen in layer 4 (Fig.~\ref{fig:results_psa_errorv_a005_L4}~(b)). This error hotspot corresponds to areas where the transient signal strength in neighbouring segments is minimal (see signal at R=22~mm in Fig.~\ref{fig:Signals_ADL_R}), reducing the intrinsic position sensitivity. This challenge particularly affects the PSA algorithms, whereas the direct NN model (panel (a)) proves more resilient, exhibiting a significantly smaller increase in error in this region.
A direct comparison between the two figures illustrates the general trend of increasing error towards the back of the crystal. The overall error in layer 6 (Fig.~\ref{fig:results_psa_errorv_a005_L6}) is visibly higher for all methods. This is attributed to two main factors: the larger segment volumes at the back reduce position sensitivity, and more importantly, the increased probability of random coincidences and scattering effects from preceding layers in the vertical scan. The performance degradation caused by these effects is significantly more pronounced in the final layer of the crystal.

\subsection{PSA Performance Relative to NN Predictions}
\label{sec:results_psa_vs_nn}

Given the potential contamination from random coincidences and mislabelled events in the scanning data, comparing PSA results directly to the scanned position might not be the most accurate way to assess the intrinsic quality of the bases. The NN model itself, trained on the large dataset, might provide a more robust reference position for each super-trace.

Therefore, to further evaluate the performance of each PSA bases, the results were compared to the positions predicted by the NN model itself. The 3D position resolution (FWHM) for each basis is summarised in Table~\ref{tab:psa_errors}.

A key finding is that the resolution values consistently improve when referenced against the NN model's predictions. For instance, in the vertical scan, the $FWHM_x$ for the ADL basis improves from 7.6~mm (vs. scan position) to 7.0~mm (vs. NN model). A similar trend holds for the NN basis, where the resolution improves from 3.1~mm to 2.5~mm.

While the resolution in the $x$-axis is largely consistent between the vertical and horizontal scans, a notable discrepancy is observed for the $y$-axis when using the simulation-based methods. This is most pronounced for the ADL basis, where the $FWHM_y$ is 7.4~mm for the vertical scan but improves significantly to 6.2~mm for the horizontal scan. A similar, though less dramatic, effect is seen for the AGATAGeFEM basis (5.5~mm vs. 5.1~mm). This discrepancy may be attributed to the different crystal illumination patterns between the two scan geometries.

These errors are significantly lower than when comparing to the scanned positions, suggesting that the PSA predictions align more closely with the NN model's output than with the ground truth from the scan position.

The analysis of PSA performance is significantly complicated by the presence of mislabelled events, such as random coincidences, which are particularly prevalent towards the back of the crystal in the vertical scan. This is evident when comparing the error maps relative to the scanned position (Fig.~\ref{fig:results_psa_errorv_a005_L6}) with those benchmarked against the NN model's predictions (Fig.~\ref{fig:results_psa_errorv_a005_nn_L6}). The substantial reduction in error in the latter case demonstrates that comparing PSA results to the NN model's output effectively neutralises the impact of these random coincidences, as both algorithms agree on the position of the outlier event, independent of the scan label.

\begin{figure*}[!t]
\centering
\includegraphics[width=1\hsize]{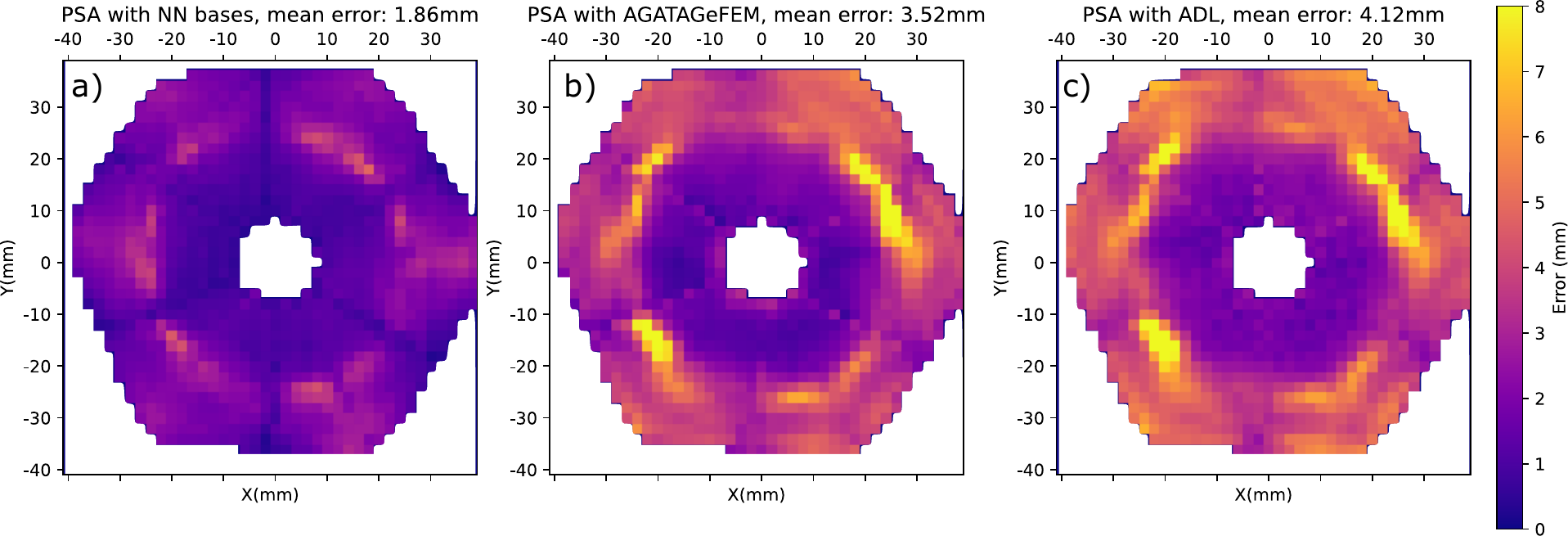}
\caption{Error map comparing PSA predictions to the direct NN model predictions for the A005 vertical scan across the fourth layer. The error is the 2D distance between the position predicted by PSA(bases) and the NN model. Panels show PSA results using:  NN bases (a), AGATAGeFEM bases (b), and ADL bases (c). Yellow colour is above 8~mm.}
\label{fig:results_psa_errorv_a005_nn_L4}
\end{figure*}

\begin{figure*}[!t]
\centering

\includegraphics[width=1\hsize]{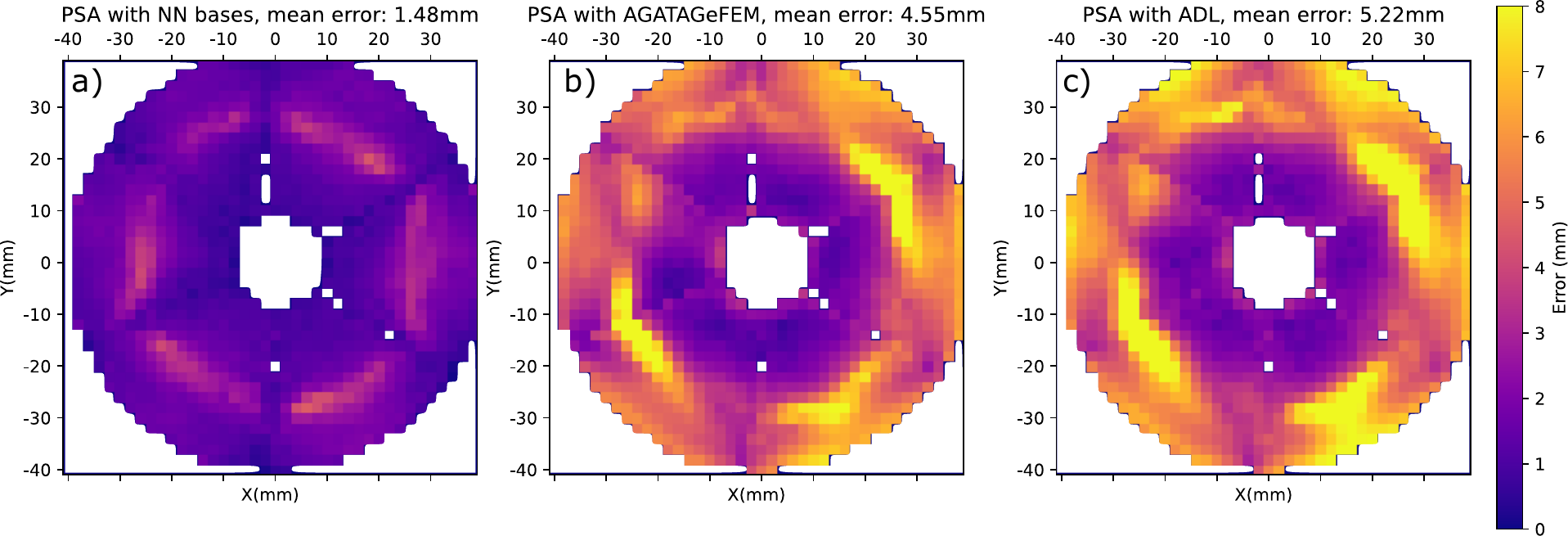} 
\caption{Error map comparing PSA predictions to NN model predictions for the sixth (back) layer. The panel arrangement is identical to the Fig.~\ref{fig:results_psa_errorv_a005_nn_L4}. Yellow colour is above 8~mm.}
\label{fig:results_psa_errorv_a005_nn_L6}
\end{figure*}

With this more robust evaluation metric established, it becomes possible to meaningfully compare the intrinsic performance of the bases across different regions of the crystal, such as the interior versus the boundaries.
Figures~\ref{fig:results_psa_errorv_a005_nn_L4} and \ref{fig:results_psa_errorv_a005_nn_L6} show this comparison for the fourth and sixth layers, respectively. In the fourth layer (Fig.~\ref{fig:results_psa_errorv_a005_nn_L4}), the self-consistency of the NN bases is immediately apparent; PSA using these bases (panel a) results in a very low, uniform error relative to the direct NN output. In contrast, the simulated AGATAGeFEM (panel b) and ADL (panel c) bases produce significantly higher and less uniform errors.
This discrepancy is even more pronounced at the crystal boundaries, as highlighted by the results from the back-most sixth layer (Fig.~\ref{fig:results_psa_errorv_a005_nn_L6}). Here, the error for the simulated bases increases substantially relative to their performance in the fourth layer. This comparison strongly suggests that while the simulated bases have general inaccuracies, these are particularly acute near the physical boundaries of the crystal, where simulations may struggle to accurately model the detector response. The experimental NN bases, however, maintain their low error even in this back layer, demonstrating their robustness across the full crystal volume.

Although this comparison indicates that the NN model provides a more robust benchmark by correctly handling outliers like random coincidences, this method has its own limitations. Specifically, it replaces the experimental uncertainty of the scan labels with the inherent, model-dependent uncertainty of the NN predictions, which is not explicitly accounted for in this analysis.

\section{Discussion}
\label{sec:discussion}

The results presented demonstrate the successful application of neural networks, specifically LSTM-based architectures, for analysing AGATA crystal scanning data obtained with the PSCS scanning method. The key achievement is the development of a method capable of predicting the full 3D interaction position from input super-traces, even when trained using data with only partial (2D) ground truth information available from the vertical and horizontal scans. This was enabled by a custom masked loss function and careful training strategies, including training separate models per segment and incorporating a Gaussian noise layer to mitigate overfitting.

The significance of this work lies primarily in the generation of high-quality experimental signal bases (NN bases) for both the S001 and A005 crystals. These bases, derived directly from experimental data using the trained NN models, led to demonstrably improved PSA performance compared to bases generated by the traditional PSCS algorithm and standard simulated bases (ADL and AGATAGeFEM). The mean PSA error was consistently lower when using NN bases indicating higher precision. Furthermore, the analysis of signal consistency using the mean standard deviation metric showed that the NN-derived signals were more homogeneous and consistent than those produced by PSCS for the S001 crystal. The comparison with simulated bases also highlighted discrepancies in the simulations, particularly regarding larger radii and potential inaccuracies near the crystal boundaries, which the NN-derived experimental bases naturally circumvent.

The experimental bases developed in this work achieve a position resolution that is nearly a factor of two better than the AGATA collaboration's 5~mm FWHM specification. This significantly surpasses the performance of the simulated bases, which, are close to the specification, but provide a substantially poorer resolution.

Despite the promising results, certain limitations should be acknowledged, while filtering methods were applied, the training data likely contained residual contamination from random coincidences, multiple-hit events, or other experimental artifacts. The robustness of the NN model against imperfections in the training data was therefore explicitly tested~\cite{Abushawish2024}. By deliberately corrupting 50\% of the position labels within a segment's dataset, it was found that the model could still achieve a prediction error close to that obtained with clean data (2.0~mm vs 1.9~mm), particularly when using the masked Euclidean distance loss function, which is inherently less sensitive to outliers than MSE. This result is significant because real experimental scanning data inevitably contains a fraction of mislabelled events. This result increases confidence that the NN model can effectively learn the correct signal-position relationships even from imperfect datasets, supporting the reliability of the generated experimental bases. This resilience mitigates concerns about data purity to some extent.

Comparing the NN approach to the established PSCS algorithm, several advantages emerge. While PSCS provides a valuable algorithmic benchmark, the NN method demonstrated superior signal consistency (lower MSD) and resulted in better PSA performance. Computationally, training the NN models requires significant resources (around 10 hours on 3 NVIDIA RTX6000 GPUs for one crystal), but the subsequent inference step (predicting positions for the entire dataset or generating bases) is relatively fast (approx. 3 hours). For comparison, the standard PSCS algorithm requires roughly 4.7 days to complete the processing on a 2.60 GHz processor machine. Once trained, applying the NN model is straightforward and readily integrated into automated analysis workflows, potentially simplifying the process compared to the potentially complex PSCS algorithm. 

An additional useful outcome of applying the NN model is the generation of a large dataset where each experimental super-trace is associated with an estimated 3D interaction position. While the original scanning data only provides partial 2D labels, and simulations might not capture all experimental details, this NN-derived dataset offers comprehensive position estimates for a large portion of the recorded events. This dataset serves two practical purposes. First, it allows for a more detailed characterisation of traditional PSA algorithm performance~\cite{Abushawish2024}. By comparing PSA results against the NN model's 3D position estimates across the detector volume (as explored in Section~\ref{sec:results_psa_vs_nn}), one can gain insights into PSA accuracy and limitations. Second, this dataset provides suitable training material for exploring machine learning models aimed directly at position prediction to replace the current PSA algorithm.

The successful validation using simulated 1D scans opens new possibilities for future characterisation campaigns. If a model can indeed be reliably trained using only one known coordinate per signal, it suggests that scanning could potentially be performed using three orthogonal 1D scans instead of two 2D scans. This could significantly reduce the time required for crystal characterisation. The concept of using 1D scanning for detector characterisation is not new and has been effectively demonstrated in the context of PET scanners using fan-beam collimators~\cite{fan_beam,fan_beam1,fan_beam4}. Interestingly, the analysis of that data often relied on machine learning models such as k-nearest neighbour algorithms or gradient-boosted decision trees. This choice of simpler models was well-suited to the nature of PET detector signals, which typically have a smaller size and less complex structure compared to the super-traces from AGATA.

Future work could extend this research in several directions. Applying the methodology to other AGATA crystals (e.g., type B or C). Exploring alternative NN architectures, such as attention mechanisms within RNNs, might yield further improvements. Further development of data cleaning and filtering techniques, potentially using ML-based anomaly detection, could enhance the quality of the training data and the resulting bases.

\section{Conclusions}
\label{sec:conclusions}

This work successfully demonstrated the development and application of a machine learning approach, utilising LSTM-based neural networks, to analyse experimental data from AGATA crystal scanning campaigns. By employing a masked loss function and tailored training strategies, the NN model effectively learned to predict 3D gamma-ray interaction positions from super-traces despite being trained on data with only 2D ground truth information.

The primary outcome is the generation of experimental signal bases for the AGATA S001 and A005 crystals directly from scanning data. These NN-derived bases exhibit high signal consistency and, when used in standard PSA algorithms, yield significantly improved position resolution compared to both the traditional PSCS algorithm and simulated bases (ADL and AGATAGeFEM). The mean PSA error was reduced, and problematic regions observed with other bases were mitigated.

This study highlights the power of machine learning as a tool for processing complex detector signals and overcoming limitations inherent in traditional analysis methods or simulations. The results suggest that NN-based approaches can provide a more accurate and potentially more efficient pathway to generating high-fidelity experimental bases crucial for optimising the performance of position-sensitive detectors like AGATA. Furthermore, the findings provide encouraging, albeit preliminary, evidence for the feasibility of potentially faster 1D scanning techniques for future detector characterisation. This work serves as a foundational step toward developing advanced machine learning-based PSA algorithms capable of resolving multiple interactions within a single segment. Overall, this research underscores the growing role of machine learning in advancing data analysis capabilities in experimental nuclear physics.

\section*{Acknowledgments}

The authors would like to thank the the Institut Pluridisciplinaire Hubert Curien (IPHC), Strasbourg, for providing the experimental scanning data and the comparative PSCS bases, and the AGATA collaboration for the availability of the S001 and A005 crystals.

\bibliographystyle{unsrt}
\bibliography{main}

@book{jrjc,
  author    = {Elsa Guy and Marie van Uffelen and Johan Loizeau and Jordan Koechler and Antoine Syx and others},
  title     = {JRJC 2022 - Journées de Rencontres Jeunes Chercheurs. Book of Proceedings},
  year      = {2023},
  note      = {⟨hal-04118967⟩},
  url       = {https://hal.science/hal-04118967}
}

@article{abadi2016tensorflow,
  title={Tensorflow: Large-scale machine learning on heterogeneous distributed systems},
  author={Abadi, Mart{\'\i}n and Agarwal, Ashish and Barham, Paul and Brevdo, Eugene and Chen, Zhifeng and Citro, Craig and Corrado, Greg S and Davis, Andy and Dean, Jeffrey and Devin, Matthieu and others},
  journal={arXiv preprint arXiv:1603.04467},
  year={2016},
  url = {https://arxiv.org/abs/1603.04467}
}

@misc{keras2015-library,
  author = {Chollet, François and others},
  title = {Keras},
  year = {2015},
  howpublished = {\url{https://keras.io}},
  note = {Software available from keras.io}
}

@misc{agapro,
  author = {IPNL GAMMA Collaboration},
  title = {AGAPRO},
  howpublished = {\url{https://gitlab.in2p3.fr/IPNL_GAMMA/narval_emulator}},
  year = {2024},
  note = {Accessed: 2024-10-03}
}

@misc{scanneddatareader,
  author = {{IP2I Gamma}},
  title = {{Scanned Data Reader}},
  howpublished = {\url{https://gitlab.in2p3.fr/ip2igamma/scanneddatareader}},
  note = {Accessed: 2024-10-05}
}

@article{nelson2007characterisation,
  title={Characterisation of an AGATA symmetric prototype detector},
  author={Nelson, L and Dimmock, Matthew Richard and Boston, AJ and Boston, HC and Cresswell, JR and Nolan, PJ and Lazarus, I and Simpson, J and Medina, P and Santos, C and others},
  journal={Nuclear Instruments and Methods in Physics Research Section A: Accelerators, Spectrometers, Detectors and Associated Equipment},
  volume={573},
  number={1-2},
  pages={153--156},
  year={2007},
  publisher={Elsevier},
  doi = {https://doi.org/10.1016/j.nima.2006.11.042}
}

@article{ha2013new,
  title={New setup for the characterisation of the AGATA detectors},
  author={Ha, TMH and Korichi, A and Le Blanc, F and D{\'e}sesquelles, P and Dosme, N and Grave, X and Karkour, N and Leboutelier, S and Legay, E and Linget, D and others},
  journal={Nuclear Instruments and Methods in Physics Research Section A: Accelerators, Spectrometers, Detectors and Associated Equipment},
  volume={697},
  pages={123--132},
  year={2013},
  publisher={Elsevier},
  doi = {https://doi.org/10.1016/j.nima.2012.08.111}
}

@article{goel2011spatial,
  title={Spatial calibration via imaging techniques of a novel scanning system for the pulse shape characterisation of position sensitive HPGe detectors},
  author={Goel, N and Domingo-Pardo, C and Engert, T and Gerl, J and Kojouharov, I and Schaffner, H},
  journal={Nuclear Instruments and Methods in Physics Research Section A: Accelerators, Spectrometers, Detectors and Associated Equipment},
  volume={652},
  number={1},
  pages={591--594},
  year={2011},
  publisher={Elsevier},
  doi = {https://doi.org/10.1016/j.nima.2011.01.146}
}

@article{hernandez2013characterization,
  title={Characterization of a High Spatial Resolution $\gamma $ Camera for Scanning HPGe Segmented Detectors},
  author={Hernandez-Prieto, A and Quintana, B},
  journal={IEEE Transactions on Nuclear Science},
  volume={60},
  number={6},
  pages={4719--4726},
  year={2013},
  publisher={IEEE},
  doi = {https://doi.org/10.1109/TNS.2013.2287252}
}

@article{PSCS0,
title = {A novel technique for the characterization of a HPGe detector response based on pulse shape comparison},
journal = {Nuclear Instruments and Methods in Physics Research Section A: Accelerators, Spectrometers, Detectors and Associated Equipment},
volume = {593},
number = {3},
pages = {440-447},
year = {2008},
issn = {0168-9002},
doi = {https://doi.org/10.1016/j.nima.2008.05.057},
url = {https://www.sciencedirect.com/science/article/pii/S0168900208007821},
author = {F.C.L. Crespi and F. Camera and B. Million and M. Sassi and O. Wieland and A. Bracco}
}

@PHDTHESIS{PSCS,
url = "http://www.theses.fr/2015STRAE047",
title = "Characterization of high-purity, multi-segmented germanium detectors",
author = "Ginsz, Michael",
year = "2015",
note = "Thèse de doctorat dirigée par Duchêne, Gilbert Physique Strasbourg 2015",
note = "2015STRAE047",
url = "http://www.theses.fr/2015STRAE047/document",
}

@PHDTHESIS{decan2020,
url = "http://www.theses.fr/2020STRAE008",
title = "3D characterization of multi-segmented HPGe detectors : simulation and validation of the PSCS technique and its application for different energies with a 152 Eu source",
author = "De Canditiis, Bartolomeo",
year = "2020",
note = "Thèse de doctorat dirigée par Duchêne, Gilbert Physique Strasbourg 2020",
note = "2020STRAE008",
url = "http://www.theses.fr/2020STRAE008/document",
}

@article{PSCS2,
  author    = {De Canditiis, B. and Duchêne, G.},
  title     = {{Simulations using the pulse shape comparison scanning technique on an AGATA segmented HPGe gamma-ray detector}},
  journal   = {The European Physical Journal A},
  year      = {2020},
  month     = {oct},
  volume    = {56},
  number    = {10},
  pages     = {276},
  issn      = {1434-601X},
  doi       = {10.1140/epja/s10050-020-00287-6},
  url       = {https://doi.org/10.1140/epja/s10050-020-00287-6}
}

@article{PSCS3,
  title        = {{Full-volume characterization of an {AGATA} segmented {HPGe} gamma-ray detector using a {$^{152}$Eu} source}},
  author       = {De Canditiis, B. and Duch{\^e}ne, G. and Sigward, M. H. and Filliger, M. and Didierjean, F. and Ginsz, M. and Ralet, D.},
  journal      = {The European Physical Journal A},
  volume       = {57},
  number       = {7},
  pages        = {223}, 
  year         = {2021},
  doi          = {10.1140/epja/s10050-021-00537-1},
  issn         = {1434-601X},
}

@article{PSA,
  title={Agata characterisation and pulse shape analysis},
  author={Boston, AJ and Crespi, FCL and Duch{\^e}ne, G and D{\'e}sesquelles, P and Gerl, J and Holloway, F and Judson, DS and Korichi, A and Harkness-Brennan, L and Ljungvall, J and others},
  journal={The European Physical Journal A},
  volume={59},
  number={9},
  pages={213},
  doi = {https://doi.org/10.1140/epja/s10050-023-01100-w},
  year={2023},
  publisher={Springer}
}

@ARTICLE{TNT_digi,
  author={Arnold, L. and Baumann, R. and Chambit, E. and Filliger, M. and Fuchs, C. and Kieber, C. and Klein, D. and Medina, P. and Parisel, C. and Richer, M. and Santos, C. and Weber, C.},
  journal={IEEE Transactions on Nuclear Science}, 
  title={TNT digital pulse processor}, 
  year={2006},
  volume={53},
  number={3},
  pages={723-728},
  doi={10.1109/TNS.2006.873712}}

@article{ML_NP,
  title={Colloquium: Machine learning in nuclear physics},
  author={Boehnlein, Amber and Diefenthaler, Markus and Sato, Nobuo and Schram, Malachi and Ziegler, Veronique and Fanelli, Cristiano and Hjorth-Jensen, Morten and Horn, Tanja and Kuchera, Michelle P and Lee, Dean and others},
  journal={Reviews of Modern Physics},
  volume={94},
  number={3},
  pages={031003},
  year={2022},
  publisher={APS},
doi={https://dx.doi.org/10.1103/RevModPhys.94.031003}
}

@article{Akkoyun2012,
  title={Agata—advanced gamma tracking array},
  author={Akkoyun, SERKAN and Algora, Alejandro and Alikhani, B and Ameil, F and De Angelis, G and Arnold, L and Astier, A and Ata{\c{c}}, Ayse and Aubert, Y and Aufranc, C and others},
  journal={Nuclear Instruments and Methods in Physics Research Section A: Accelerators, Spectrometers, Detectors and Associated Equipment},
  volume={668},
  pages={26--58},
  year={2012},
  publisher={Elsevier},
  url= "https://doi.org/10.1016/j.nima.2011.11.081"
}

@article{AGATAGeFEM,
  author = {J. Ljungvall},
  title = {Pulse-shape calculations and applications using the AGATAGeFEM software package},
  journal = {The European Physical Journal A},
  volume = {57},
  number = {6},
  pages = {198},
  year = {2021},
  doi = {10.1140/epja/s10050-021-00512-w},
  url = {https://www.agata.org/node/445}
}

@article{kingma2014adam,
  title={Adam: A method for stochastic optimization},
  author={Kingma, Diederik P. and Ba, Jimmy},
  journal={arXiv preprint arXiv:1412.6980},
  year={2014},
  url={https://arxiv.org/abs/1412.6980}
}

@article{ADL,
  author = {Bruyneel, B. and Birkenbach, B. and Reiter, P.},
  title = {Pulse shape analysis and position determination in segmented HPGe detectors: The AGATA detector library},
  journal = {The European Physical Journal A},
  volume = {52},
  number = {3},
  pages = {70},
  year = {2016},
  doi = {10.1140/epja/i2016-16070-9},
  url = {https://doi.org/10.1140/epja/i2016-16070-9},
  issn = {1434-601X},
  month = {03}
}

@article{GANIL_AGATA,
title = {{Conceptual design of the AGATA $\pi$ array at GANIL}},
journal = {Nuclear Instruments and Methods in Physics Research Section A: Accelerators, Spectrometers, Detectors and Associated Equipment},
volume = {855},
pages = {1-12},
year = {2017},
issn = {0168-9002},
doi = {https://doi.org/10.1016/j.nima.2017.02.063},
url = {https://www.sciencedirect.com/science/article/pii/S0168900217302590},
author = {E. Clément and C. Michelagnoli and G. {de France} and H.J. Li and A. Lemasson and C. {Barthe Dejean} and M. Beuzard and P. Bougault and J. Cacitti and J.-L. Foucher and G. Fremont and P. Gangnant and J. Goupil and C. Houarner and M. Jean and A. Lefevre and L. Legeard and F. Legruel and C. Maugeais and L. Ménager and N. Ménard and H. Munoz and M. Ozille and B. Raine and J.A. Ropert and F. Saillant and C. Spitaels and M. Tripon and Ph. Vallerand and G. Voltolini and W. Korten and M.-D. Salsac and Ch. Theisen and M. Zielińska and T. Joannem and M. Karolak and M. Kebbiri and A. Lotode and R. Touzery and Ch. Walter and A. Korichi and J. Ljungvall and A. Lopez-Martens and D. Ralet and N. Dosme and X. Grave and N. Karkour and X. Lafay and E. Legay and I. Kojouharov and C. Domingo-Pardo and A. Gadea and R.M. Pérez-Vidal and J.V. Civera and B. Birkenbach and J. Eberth and H. Hess and L. Lewandowski and P. Reiter and A. Nannini and G. {De Angelis} and G. Jaworski and P. John and D.R. Napoli and J.J. Valiente-Dobón and D. Barrientos and D. Bortolato and G. Benzoni and A. Bracco and S. Brambilla and F. Camera and F.C.L. Crespi and S. Leoni and B. Million and A. Pullia and O. Wieland and D. Bazzacco and S.M. Lenzi and S. Lunardi and R. Menegazzo and D. Mengoni and F. Recchia and M. Bellato and R. Isocrate and F.J. {Egea Canet} and F. Didierjean and G. Duchêne and R. Baumann and M. Brucker and E. Dangelser and M. Filliger and H. Friedmann and G. Gaudiot and J.-N. Grapton and H. Kocher and C. Mathieu and M.-H. Sigward and D. Thomas and S. Veeramootoo and J. Dudouet and O. Stézowski and C. Aufranc and Y. Aubert and M. Labiche and J. Simpson and I. Burrows and P.J. Coleman-Smith and A. Grant and I.H. Lazarus and P.S. Morrall and V.F.E. Pucknell and A. Boston and D.S. Judson and N. Lalović and J. Nyberg and J. Collado and V. González and I. Kuti and B.M. Nyakó and A. Maj and M. Rudigier}
}

@article{PSA2,
author = {R. Venturelli and D. Bazzacco},
journal = {LNL Annual Report 2004},
title = {{Adaptive Grid Search as Pulse Shape Analysis Algorithm for $\gamma$-Tracking and Results}},
year = {2004},
url = {http://www.micros2005.lnl.infn.it/~annrep/read_ar/2004/contrib_2004/pdfs/FAA122.pdf}
}

@article{AGATA2,
title = {The Advanced Gamma Ray Tracking Array AGATA},
journal = {Nuclear Physics A},
volume = {746},
pages = {248-254},
year = {2004},
note = {Proceedings of the Sixth International Conference on Radioactive Nuclear Beams (RNB6)},
issn = {0375-9474},
doi = {https://doi.org/10.1016/j.nuclphysa.2004.09.148},
url = {https://www.sciencedirect.com/science/article/pii/S0375947404009625},
author = {Dino Bazzacco}
}

@Inbook{LSTM,
author="Graves, Alex",
title="Long Short-Term Memory",
bookTitle="Supervised Sequence Labelling with Recurrent Neural Networks",
year="2012",
publisher="Springer Berlin Heidelberg",
address="Berlin, Heidelberg",
pages="37--45",
isbn="978-3-642-24797-2",
doi="10.1007/978-3-642-24797-2_4",
url="https://doi.org/10.1007/978-3-642-24797-2_4"
}

@article{FABIAN,
title = {{Artificial neural networks for neutron/$\gamma$ discrimination in the neutron detectors of NEDA}},
journal = {Nuclear Instruments and Methods in Physics Research Section A: Accelerators, Spectrometers, Detectors and Associated Equipment},
volume = {986},
pages = {164750},
year = {2021},
issn = {0168-9002},
doi = {https://doi.org/10.1016/j.nima.2020.164750},
url = {https://www.sciencedirect.com/science/article/pii/S0168900220311475},
author = {X. Fabian and G. Baulieu and L. Ducroux and O. Stézowski and A. Boujrad and E. Clément and S. Coudert and G. {de France} and N. Erduran and S. Ertürk and V. González and G. Jaworski and J. Nyberg and D. Ralet and E. Sanchis and R. Wadsworth}
}

@software{uproot,
  author       = {Pivarski, Jim and
                  Schreiner, Henry and
                  Hollands, Angus and
                  Das, Pratyush and
                  Kothari, Kush and
                  Roy, Aryan and
                  Ling, Jerry and
                  Smith, Nicholas and
                  Burr, Chris and
                  Stark, Giordon},
  title        = {Uproot},
  month        = aug,
  year         = 2023,
  publisher    = {Zenodo},
  version      = {v5.0.11},
  doi          = {10.5281/zenodo.8239801},
  url          = {https://doi.org/10.5281/zenodo.8239801},
}

@article{stez,
  title        = {{Advancements in software developments}},
  author       = {St{\'{e}}zowski, O. and Dudouet, J. and Goasduff, A. and Korichi, A. and Aubert, Y. and Balogh, M. and Baulieu, G. and Bazzacco, D. and Brambilla, S. and Brugnara, D. and Dosme, N. and Elloumi, S. and Gauron, P. and Grave, X. and Jacob, J. and Lafage, V. and Lemasson, A. and Legay, E. and Le Jeannic, P. and Ljungvall, J. and Matta, A. and Molina, R. and Philippon, G. and Sedlak, M. and Taurigna-Quere, M. and Toniolo, N.},
  journal      = {The European Physical Journal A},
  volume       = {59},
  number       = {5},
  pages        = {119},
  year         = {2023},
  doi          = {10.1140/epja/s10050-023-01025-4},
  issn         = {1434-601X},
}

@article{PSA3,
title = {Position resolution of the prototype AGATA triple-cluster detector from an in-beam experiment},
journal = {Nuclear Instruments and Methods in Physics Research Section A: Accelerators, Spectrometers, Detectors and Associated Equipment},
volume = {604},
number = {3},
pages = {555-562},
year = {2009},
issn = {0168-9002},
doi = {https://doi.org/10.1016/j.nima.2009.02.042},
url = {https://www.sciencedirect.com/science/article/pii/S0168900209004124},
author = {F. Recchia and D. Bazzacco and E. Farnea and A. Gadea and R. Venturelli and T. Beck and P. Bednarczyk and A. Buerger and A. Dewald and M. Dimmock and G. Duchêne and J. Eberth and T. Faul and J. Gerl and R. Gernhaeuser and K. Hauschild and A. Holler and P. Jones and W. Korten and Th. Kröll and R. Krücken and N. Kurz and J. Ljungvall and S. Lunardi and P. Maierbeck and D. Mengoni and J. Nyberg and L. Nelson and G. Pascovici and P. Reiter and H. Schaffner and M. Schlarb and T. Steinhardt and O. Thelen and C.A. Ur and J.J. {Valiente Dobon} and D. Weißhaar}
}

@phdthesis{Abushawish2024,
TITLE = {{New Machine Learning-Based Approaches for AGATA Detectors Characterization and Nuclear Structure Studies of Neutron-Rich Nb Isotopes}},
AUTHOR = {Abushawish, Mojahed},
URL = {https://theses.hal.science/tel-04966124},
NUMBER = {2024LYO10344},
SCHOOL = {{Universit{\'e} Claude Bernard - Lyon I}},
YEAR = {2024},
MONTH = Dec,
TYPE = {Theses},
HAL_ID = {tel-04966124},
HAL_VERSION = {v1},
}

@ARTICLE{fan_beam,
  author={van Dam, Herman T. and Seifert, Stefan and Vinke, Ruud and Dendooven, Peter and Lohner, Herbert and Beekman, Freek J. and Schaart, Dennis R.},
  journal={IEEE Transactions on Nuclear Science}, 
  title={Improved Nearest Neighbor Methods for Gamma Photon Interaction Position Determination in Monolithic Scintillator PET Detectors}, 
  year={2011},
  volume={58},
  number={5},
  pages={2139-2147},
  doi={10.1109/TNS.2011.2150762}}

@article{fan_beam4,
author = {Kuhl, Yannick and Mueller, Florian and Thull, Julian and Naunheim, Stephan and Schug, David and Schulz, Volkmar},
title = {3D in-system calibration method for PET detectors},
journal = {Medical Physics},
volume = {52},
number = {1},
pages = {232-245},
doi = {https://doi.org/10.1002/mp.17475},
url = {https://aapm.onlinelibrary.wiley.com/doi/abs/10.1002/mp.17475},
eprint = {https://aapm.onlinelibrary.wiley.com/doi/pdf/10.1002/mp.17475},
year = {2025}
}

@ARTICLE{fan_beam1,
  author={Müller, Florian and Schug, David and Hallen, Patrick and Grahe, Jan and Schulz, Volkmar},
  journal={IEEE Transactions on Radiation and Plasma Medical Sciences}, 
  title={Gradient Tree Boosting-Based Positioning Method for Monolithic Scintillator Crystals in Positron Emission Tomography}, 
  year={2018},
  volume={2},
  number={5},
  pages={411-421},
  doi={10.1109/TRPMS.2018.2837738}}

@article{S001,
title = {Characterisation of an AGATA symmetric prototype detector},
journal = {Nuclear Instruments and Methods in Physics Research Section A: Accelerators, Spectrometers, Detectors and Associated Equipment},
volume = {573},
number = {1},
pages = {153-156},
year = {2007},
note = {Proceedings of the 7th International Conference on Position-Sensitive Detectors},
issn = {0168-9002},
doi = {https://doi.org/10.1016/j.nima.2006.11.042},
url = {https://www.sciencedirect.com/science/article/pii/S0168900206022479},
author = {L. Nelson and M.R. Dimmock and A.J. Boston and H.C. Boston and J.R. Cresswell and P.J. Nolan and I. Lazarus and J. Simpson and P. Medina and C. Santos and C. Parisel}
}

\end{document}